# A full contraction-reaction-diffusion model for pattern formation in geometrically confined microtissues

Tiankai Zhao[1], Hongyan Yuan[1*]

[1]Department of Mechanics and Aerospace Engineering, Southern University of Science and Technology, Shenzhen, Guangdong 518055, China

*Corresponding Author: yuanhy3@sustech.edu.cn

**Abstract**
The reaction-diffusion models have been extensively applied to explain the mechanism of pattern formations in early embryogenesis based on geometrically confined microtissues consisting of human pluripotent stem cells. Recently, mechanical cues, such as the cellular stresses and strains, have been found to dictate the pattern formation in human stem cell differentiation. As a result, the traditional reaction-diffusion models are modified by adding mechanically related terms to consider the role played by the mechanical cues. However, these models either do not consider the activeness of the cellular tissues or neglect their poroelastic nature that biological tissues are made by both cells and interstitial fluid. Hence, the current models suffer from the lacks of biophysical relevance. Here we propose a modified reaction-diffusion model that couples with the active contraction of cellular tissues. The cellular tissue is modelled as a piece of biphasic poroelastic material, where mechanical forces naturally regulate the transport of chemical cues. Such chemical cues direct cell fate and hence yield certain types of pattern formations observed in previous experiments.

## 1. Introduction
Pattern formations based on the reaction-diffusion of morphogens in developmental biology have drawn scientists' interest for quite a long time[1–4]. Pairs of morphogens, called the activator and inhibitor, are proposed to be responsible for the pattern formation in stem cell differentiation during embryogenesis[1,2]. The concepts of morphogens and their interactions have been long proposed, and during decades people have applied them onto the explanations of pattern formation phenomena[5], such as the determination of the hair follicle space by Wnt and Dkk[6], the regulation of stem cell activation during hair regeneration by FGF (fibroblast growth factor) and BMP (bone morphogenetic protein)[7], and Rho signaling (as activators) and F-actin assembly (as inhibitors) during animal cell cytokinesis[8]. In these processes, the interactions between the activators and inhibitors are usually complicated[8,9], but sometimes it can be summarized in a simple scheme. That is the presence of the activator usually promotes both the production of itself and the inhibitor; the inhibitor, in tun, will prevent the generation of the activator[4], as shown in Fig. 1(b). Such interactions can be mathematically described by several models, including Gierer-Meinhardt model[2], Lengyel-Epstein model[10], and Barrio-Varea-Aragon-Maini (BVAM) model[11]. Once secreted by cells through exocytosis[12], the morphogens will be transported in the interstitial fluid in-between cells by advection and diffusion, which is hindered by the tortuosity of cells[13]. The transported morphogens are then uptaken by neighboring cells through endocytosis[14,15] or act through cell surface receptors[16,17]. In such ways, morphogens can collectively coordinate the fate of a group of cells.

The reaction-diffusion model has also been applied to explain the pattern formation in human pluripotent stem cells (hPSCs)-based in vitro models for the study of early-stage human development[5,18,19]. In a typical hPSC-based model, hPSCs were cultured in mesoscale patterns with different geometries, such as circles[20] and triangles[21]. A uniform distribution of morphogens is created in the culture media of the cells. Under such conditions, the hPSC-based cell colony with a circular geometry will differentiate into self-organized concentric rings of different cell types, mimicking the three-germ-layer formation during gastrulation[15,19], or neuroepithelium/neural crest/epidermis patterning during neural plate induction[23–25]. The former is regulated by a reaction-diffusion process of BMP4 and its inhibitor Noggin[18], while the latter is modulated by the reaction-diffusion of BMP-Noggin and Wnt-Dkk pairs[25]. In these works, axisymmetric reaction-diffusion equations, along with an axisymmetric initial condition are directly imposed[5,18,25], as normal reaction-diffusion models cannot predict a concentric-ring pattern of differentiated cells[19]. Until very recently, a contraction-reaction-diffusion model[19] has been introduced to explain the ring pattern formation by emphasizing the role played by the active forces generated in cells.

Mechanics-guided spatial-temporal evolutions of bio-molecules have been extensively studied, such as the formation of receptor-ligand complexes in focal adhesion formation[26,27], principal-stress-direction-guided myofibril organizations in cell adhesion and migration[28–31], intracellular-stress-dictated myosin distribution during cell polarization[32], and the morphological evolution under tissue level regulated by mechanochemo-coupling process[33,34]. For the pattern formation during early embryogenesis, experimental studies reveal that mechanical cues could also play a dominating role in dictating the distribution of signaling molecules[21,24,35]. Based on these findings, quite a number of mechanical-coupling reaction-diffusion models have been proposed to study the pattern formation in stem cell differentiation during early embryogenesis[19,36–39]. In these studies, the cellular tissues have been modelled as homogeneous single-phase elastic solids, where the morphogens/signaling molecules can diffuse across the whole piece of the material. Yet they neglect the fact that the diffusion can only happen in the interstitial fluid in-between cells[13,40], which is due to the porous nature of biological tissues[41]. Additionally, the cellular and tissue properties, such as the active contractions, can also be affected by the morphogens, which is not considered by these works. During recent decades, quite a few models have considered the tissues as porous materials when studying the reaction-diffusion of morphogens. Dhote and Vernerey[42] studied the degradation of a porous hydrogel scaffold caused by the diffusion of enzyme (like MMPs) during early stage of tissue growth. Armstrong et al.[43] proposed a model for mixed porohyperelasticity with transport, swelling, and growth on the development of artery. De Oliveira Vilaca et al.[44] considered the brain tissue as poroelastic material to obtain the stress-strain field within the tissue and consider it as an input source term for the diffusion of interacting solutes. These models, however, mainly focus on the morphological evolution of organs during the late stages of embryo development. Recently, Recho et al.[45] have modeled the cellular tissue as a biphasic porous material where morphogens are secreted by cells and can only diffuse in the interstitial fluid. Yet, this model has not considered the active contraction of the tissue and made a lot of mathematical assumptions to keep the model complexity at its minimum. And how the spatial-temporal distribution of the morphogens affects the mechanics of the cellular tissues is also out of its considerations. Thus, a full contraction-reaction-diffusion model with the consideration of the porous nature of biological tissues needs to be developed for pattern formations during very early embryogenesis.

In this article, we have developed a two-dimensional contraction-reaction-diffusion model by considering the cellular tissue, such as the patterned hPSCs colony, being biphasic poroelastic material. The active contraction of cells on substrate has been included in the model. The whole tissue is modeled as a biphasic material with a cellular phase and an interstitial fluid phase. The chemical reactions of the morphogens only take place within the cellular phase. The generated morphogens are then secreted from cells into the interstitial fluid by exocytosis. Enlightened by the previous study on the morphogen gradient formation in two-dimensional biological tissues[46], we hypothesize that the morphogens are mainly transported in the interstitial fluid though pressure-induced advection and diffusion, and are uptaken by cells through endocytosis. The resulted spatial-temporal distribution of the morphogens, in turn, can shape the evolutions of the cellular phase as well as the tissue active contractility. We note that our model can naturally take the effect of the mechanical stresses into account and successfully demonstrate the ring pattern formation in hPSCs-based in-virto model.

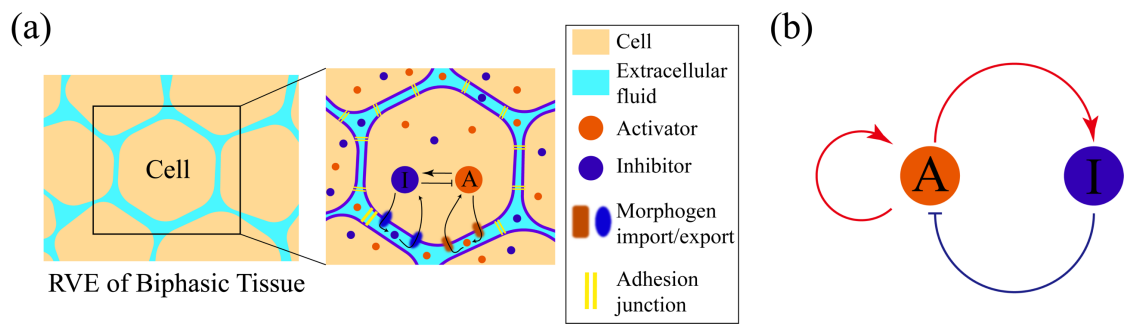

**Fig. 1.** Illustrations of the model for the pattern formation in the two-dimensional biphasic porous tissues. (a) Schematic of the representative volume element (RVE) of the model. Cells form a poroelastic network, permeated by interstitial fluid. The biochemical interactions between morphogens, $A$, the activator and $I$, the inhibitor, take place inside the cell and are exported across the cell membrane trough endocytosis and exocytosis. They are transported in the fluid by advection and diffusion. (b) Schematic of the biochemical interactions between $A$ and $I$. The activator promotes the generation of both itself and the inhibitor; the inhibitor prevents the generation of the activator.

## 2. Theory
*2.1 Modeling of the two-dimensional developmental tissue as biphasic materials*

A developmental multicellular tissue which is much larger than its individual cells can be modeled as a piece of active biphasic porous material, with a first phase consisting of adhesive cells, interconnected by cell-cell junctions and a second phase made of extracellular fluid. The tissue is patterned on substrate, forming a cohesive cell monolayer that occupies an area $\Omega$. As sketched in Fig. 1 (a), the representative volume element (RVE) with a uniform thickness $h_c$ is occupied by cells with random shapes, whose area fraction is averaged in the RVE and is denoted by the phase variable $\phi_c(\mathbf{x}, t)$, satisfying $0 \leq \phi_c \leq 1$. The rest part of the RVE is filled of interstitial fluid flowing in-between cells, whose area fraction is denoted by the phase variable $\phi_f(\mathbf{x}, t)$. The two area fractions obey the law of local conservation:

$$\phi_c + \phi_f = 1. \tag{1}$$

As matters can be exchanged between cells and extracellular medium by fluid flows, the area fractions of cells and the interstitial fluid evolve with respect to time. The evolution of $\phi_c(\mathbf{x}, t)$ and $\phi_f(\mathbf{x}, t)$ satisfies the following mass balance relations:

$$\partial_t \phi_c + \nabla \cdot (\phi_c \mathbf{v}_c) = S \text{ in } \Omega, \tag{2a}$$

$$\partial_t \phi_f + \nabla \cdot (\phi_f \mathbf{v}_f) = -S \text{ in } \Omega, \tag{2b}$$

where $\mathbf{v}_{c,f}$ are in-plane fluid velocities with respect to the intracellular and extracellular flows; $S$ denotes for the exchanging rate of matters between the two phases. By summing Eq. (2a) with Eq. (2b), and noting that $\partial_t(\phi_c + \phi_f) = 0$ from Eq. (1), one can obtain the following relation between the two velocities:

$$\nabla \cdot (\phi_c \mathbf{v}_c) = -\nabla \cdot [(1 - \phi_c)\mathbf{v}_f]. \tag{3}$$

The boundary conditions for Eq. (2a) and (2b) are proposed that there is no matter flowing out of the tissue, which is to say:

$$\nabla(\phi_c \mathbf{v}_c) \cdot \mathbf{n} = 0 \text{ at } \partial\Omega, \tag{4a}$$

$$\nabla(\phi_f \mathbf{v}_f) \cdot \mathbf{n} = 0 \text{ at } \partial\Omega, \tag{4b}$$

where $\partial\Omega$ is the boundary of the region occupied by the tissue with $\mathbf{n}$ being its outer unit normal. By considering that the two-dimensional cell wants to regulate its relative area to a homeostatic value $\phi_h$, the exchanging term $S$ can be written as[45]:

$$S = \frac{\phi_h - \phi_c}{\tau}, \tag{5}$$

where $\tau$ is the time scale that characterizes the cell volume regulation rate. The homeostatic value $\phi_h$ depends on many factors, such as the active rearrangement of cell cortex[47,48], water flow caused by the osmotic pressure[47,49], and the flux of ions through transmembrane channels[50]. In this work, we focus on the situation that $\phi_h$ is 5/30/22 7:49:00 PMmainly influenced by the active levels of morphogens while the effect brought by other factors is small. The biological basis of such assumption is that morphogens can either actively control the cell volume[51] or act as growth factors/inhibitors that regulate the cell proliferation and loss[52,53]. Then $\phi_h$ is written as[45]:

$$\phi_h(A_i, I_i) = \phi^* + k_A(A_i - A_i^*) + k_I(I_i - I_i^*), \tag{6}$$

where $\phi^*$ is assumed to be a constant; $A_i(\mathbf{x}, t)$ and $I_i(\mathbf{x}, t)$ stand for the active levels of activators and inhibitors, which are characterized by their normalized relative concentrations, respectively. The activators are the morphogens that favor the cell differentiation, while the inhibitors are the morphogens that suppress this process. Together, they dictate the pattern formation during cell differentiation. Equation (6) can be viewed as a linear expansion of $\phi_h(A_i, I_i)$ with respect to $A_i$ and $I_i$ at their homeostatic values: $A_i^*$ and $I_i^*$, with $k_A = \left.\frac{\partial \phi_h}{\partial A_i}\right|_{(A_i^*, I_i^*)}$ and $k_I = \left.\frac{\partial \phi_h}{\partial I_i}\right|_{(A_i^*, I_i^*)}$.

*2.2 The kinetics of morphogens in developmental multicellular tissues*

Morphogens are molecules that are secreted by cells and travel through interstitial fluid by advection and diffusion. They serve as an intercellular communication tool and collectively determine cell fate. Most biochemical interactions of morphogens happen inside cells or on cell membranes, therefore, the evolution of the active levels of the intracellular morphogen $A_i$ and $I_i$ is associated with the cellular phase variable $\phi_c$, written as:

$$\partial_t(\phi_c A_i) = \phi_c f(A_i, I_i) + \gamma_A(A_e - K_A A_i) \text{ in } \Omega, \tag{7a}$$
$$\partial_t(\phi_c I_i) = \phi_c g(A_i, I_i) + \gamma_I(I_e - K_I I_i) \text{ in } \Omega, \tag{7b}$$

where $A_e(\mathbf{x}, t)$ and $I_e(\mathbf{x}, t)$ are the active levels of the activators and inhibitors in the interstitial fluid; $\gamma_{A,I}$ are the corresponding endocytosis rates of the two morphogens and are interpolated as: $\gamma_{A,I} = \gamma_{A,I}^0 \phi_c \phi_f$, which means there is no endocytosis when either phase is saturated (equivalent to either phase variable being zero); $K_A$ and $K_I$ are the ratios of exocytosis rates to the endocytosis rates with respect to the activators and inhibitors. The nonlinear terms $f(A_i, I_i)$ and $g(A_i, I_i)$ depict the chemical interactions between activators and inhibitors. Such typical interactions, like BMP-Noggin and Wnt-Dkk interaction pairs, can be described by Barrio-Varea-Aragon-Maini (BVAM) model[11]:

$$f(A_i, I_i) = f_{A_i^*}(A_i - A_i^*) + f_{I_i^*}(I_i - I_i^*) + f_{A_i^* I_i^* I_i^*}(A_i - A_i^*)(I_i - I_i^*)^2, \tag{8a}$$

$$g(A_i, I_i) = g_{A_i^*}(A_i - A_i^*) + g(I_i - I_i^*) + g_{A_i^* I_i^* I_i^*}(A_i - A_i^*)(I_i - I_i^*)^2, \tag{8b}$$

where the linear terms in Eq. (8a) and (8b) represent the reaction rates of the activators and inhibitors; the nonlinear terms in the first and the second equations describe the inhibitions imposed on the activators by the inhibitors, and the promotion of the activators imposed on the inhibitors, respectively[19,54].

Unlike the kinetics of the morphogens inside the cells is governed by chemical reactions along with endocytosis and exocytosis, the mass conservation of the morphogens in the interstitial fluid is dictated by advection-diffusion, and co-evolves with the interstitial fluid phase $\phi_f$, which says:

$$\partial_t(\phi_f A_e) + \nabla \cdot \mathbf{J}_A = -\gamma_A(A_e - K_A A_i) \text{ in } \Omega, \tag{9a}$$

$$\partial_t(\phi_f I_e) + \nabla \cdot \mathbf{J}_I = -\gamma_I(I_e - K_I I_i) \text{ in } \Omega, \tag{9b}$$

where the fluxes $\mathbf{J}_A$ and $\mathbf{J}_I$ contain both the advective part and diffusive part[55]:

$$\mathbf{J}_A = \phi_f A_e \mathbf{v}_f - D\phi_f \nabla A_e, \tag{10a}$$

$$\mathbf{J}_I = \phi_f I_e \mathbf{v}_f - D\phi_f \nabla I_e. \tag{10b}$$

The parameter $D$ is the global Fickian diffusivity associated with $A_e$ and $I_e$. We note that $D$ could be a function of the deformation of the multicellular tissue[56] or the cellular phase variable $\phi_c$[57]. Yet here, we adopt the assumption that $D$ is neither affected by tissue deformation nor by the cellular phase variable $\phi_c$. The former assumption holds well for small deformation cases; the latter one, on the other hand, is justified by the small change in the value of $\phi_c$ revealed by both the experimental observations on the developmental hPSC tissues in virto[24,25] and our modeling results (which will be shown in Section 5). By these two assumptions, $D$ can be considered constant within the tissue. The boundary conditions for Eq. (10a) and (10b) are proposed as:

$$\mathbf{J}_A \cdot \mathbf{n} = 0 \text{ at } \partial\Omega, \tag{11a}$$

$$\mathbf{J}_I \cdot \mathbf{n} = 0 \text{ at } \partial\Omega, \tag{11b}$$

which indicate that there is no morphogen flowing out of the developmental tissue.

*2.3 The constitutive law of the developmental tissue as an active biphasic porous material*

Previous experiments show no strong anisotropy of multicellular developmental tissues that are patterned on substrate in virto, and the small deformation assumption usually hold for such cases. Therefore, we can adopt a linear isotropic constitutive relation for the tissues based on the theory of poroelasticity. The three-dimensional linear isotropic constitutive law for porous materials says[45,58]:

$$\boldsymbol{\sigma} = 2G\boldsymbol{\varepsilon} + \left(K_u - \frac{2G}{3}\right) tr(\boldsymbol{\varepsilon})\mathbf{I} + \frac{K_u - K}{\alpha}(\phi_c - \phi^*)\mathbf{I}, \tag{12a}$$

$$p = -\frac{K_u - K}{\alpha} tr(\boldsymbol{\varepsilon}) - \frac{K_u - K}{\alpha^2}(\phi_c - \phi^*), \tag{12b}$$

where $\boldsymbol{\sigma}$ is the three-dimensional Cauchy stress tensor; $\boldsymbol{\varepsilon}$ is the corresponding small-deformation strain tensor; $\mathbf{I}$ stands for the three-by-three identity tensor; and $p$ denotes the pressure. The material parameters $K_u$ and $K$ correspond to the undrained and drained bulk moduli of the porous material, while $G$ is the shear modulus and $\alpha$ is Biot coefficient. The variables of $\phi_c$ and $\phi^*$ in Eq. (12) describe the volume fraction of cells in the current and the reference states, respectively. For a piece of hPSC developmental tissue patterned on the two-dimensional substrate, the normal stress

component in the third direction can be neglected due to its thin monolayer geometry, which reads:

$$\sigma_{33} = 0. \tag{13}$$

By substituting Eq. (11) into Eq. (10a), one can obtain:

$$\varepsilon_{33} = -\frac{\nu_u}{1-\nu_u}[\varepsilon_{11} + \varepsilon_{22}] - \frac{1-2\nu_u}{1-\nu_u}\frac{K_u-K}{2\alpha G}(\phi_c - \phi^*), \tag{14}$$

where $\nu_u$ is the undrained Poisson's ratio. By substituting Eq. (14) back into Eq. (12a) and (12b), we could derive the plane-stress linear isotropic constitutive law for porous materials that relates the two-dimensional Cauchy stress tensor $\boldsymbol{\sigma}$ and the in-plane pressure $p$ with the two-dimensional strain tensor $\boldsymbol{\varepsilon}$ and the two-by-two identity $\mathbf{I}$:

$$\boldsymbol{\sigma} = 2G\left[\boldsymbol{\varepsilon} + \frac{\nu_u}{1-\nu_u}tr(\boldsymbol{\varepsilon})\mathbf{I}\right] + \frac{1-2\nu_u}{1-\nu_u}\frac{K_u-K}{\alpha}(\phi_c - \phi^*)\mathbf{I}, \tag{15a}$$

$$p = -\frac{1-2\nu_u}{1-\nu_u}\frac{K_u-K}{\alpha}tr(\boldsymbol{\varepsilon}) - \gamma\frac{K_u-K}{\alpha^2}(\phi_c - \phi^*), \tag{15b}$$

with $\gamma = \frac{(1-2\nu_u)(1-\nu)}{(1-\nu_u)(1-2\nu)}$. The symbols $\boldsymbol{\sigma}, \boldsymbol{\varepsilon},$ and $\mathbf{I}$, without any specific notation, now stand for the variables that describe the corresponding physical quantities in the plane-stress case. The variables $\phi_c$ and $\phi^*$ describe the area fraction of cells within the two-dimensional cellular sheet, same as what they do in Section 2.1. The activeness of the two-dimensional developmental tissue can be modeled by introducing an inelastic strain tensor[27]:

$$\boldsymbol{\varepsilon}^A = -\varepsilon^A \mathbf{I}, \tag{16}$$

which characterizes the active in-plane contraction of the tissue as a whole with $\varepsilon^A > 0$. By letting $\boldsymbol{\varepsilon}^{el} = \boldsymbol{\varepsilon} - \boldsymbol{\varepsilon}^A$, the constitutive relations in Eq. (15a) and (15b) can modified to become[49]:

$$\boldsymbol{\sigma} = 2G\left[\boldsymbol{\varepsilon}^{el} + \frac{\nu_u}{1-\nu_u}tr(\boldsymbol{\varepsilon}^{el})\mathbf{I}\right] + \frac{1-2\nu_u}{1-\nu_u}\frac{K_u-K}{\alpha}[(\phi_c - \phi^*) + \beta tr(\boldsymbol{\varepsilon}^A)]\mathbf{I}, \tag{17a}$$

$$p = -\frac{1-2\nu_u}{1-\nu_u}\frac{K_u-K}{\alpha}tr(\boldsymbol{\varepsilon}^{el}) - \gamma\frac{K_u-K}{\alpha^2}[(\phi_c - \phi^*) + \beta tr(\boldsymbol{\varepsilon}^A)], \tag{17b}$$

Where $\beta$ is a material parameter and $1 - \phi^* \leq \beta \leq 1$. As the morphogens are able to dictate cell fates by differentiating cells into specific kinds, it is reasonable to hypothesize that the active contractility is also influenced by the active levels of morphogens,[44] which is $\varepsilon^A = \varepsilon_0^A \psi(A_i, I_i)$, with $\psi(A_i, I_i)$ being a function on $A_i$ and $I_i$. The two-dimensional biphasic porous tissue actively interacts with the substrate by exerting traction forces through its focal adhesions. The traction forces, which in turn are transmitted back into the tissue, are balanced with the intercellular stress through cell-cell adhesions. Therefore, the stress equilibrium and the corresponding boundary condition simply says[27,60]:

$$\nabla \cdot \boldsymbol{\sigma} - \frac{Y\phi_c}{h_c}\mathbf{u} = 0 \text{ in } \Omega, \tag{18a}$$

$$\boldsymbol{\sigma}\mathbf{n} = \mathbf{0} \text{ at } \partial\Omega, \tag{18b}$$

where $Y\phi_c$ is the effective strength of the focal adhesion which depends on area fraction of the cell. Here for simplicity, $Y$ is assumed to be homogeneous over the whole colony. The vector $\mathbf{n}$ stands for the outer unit normal of the tissue boundary.

For the case that the tissue thickness $h_c$ is much smaller than its in-plane size, the traction force per unit area $\mathbf{T} = -Y\phi_c\mathbf{u}$ can be averaged along the colony thickness and considered as a body force. Although it is not always the case, here we assume that the cellular tissue is cohesive so that the expression of the traction force follows the form in Eq. (18a).[60]

In poroelastic materials, the gradient of pressure drives the flow of the interstitial fluid, and therefore advects the morphogens. Such mechanism is simply described by Darcy's law:

$$\phi_f \mathbf{v}_f = -\frac{\kappa}{\eta}\nabla p, \tag{19}$$

where $\kappa$ is the permeability of the tissue; $\eta$ is the fluid viscosity. By substituting Eq. (19) into Eq. (2a), Eq. (3), Eq. (4a), and Eq. (5), one can obtain the governing equation and associated boundary condition for $\phi_c$:

$$\partial_t \phi_c + \nabla \cdot \left(\frac{\kappa}{\eta}\nabla p\right) = \frac{\phi_h - \phi_c}{\tau} \quad \text{in } \Omega, \tag{20a}$$

$$\frac{\kappa}{\eta}\nabla p \cdot \mathbf{n} = 0 \quad \text{at } \partial\Omega. \tag{20b}$$

Substituting Eq. (19) into Eq. (9), Eq. (10), and Eq. (11), one can derive the governing equations and associated boundary conditions for $A_e$ and $I_e$ in a compact form:

$$\partial_t(\phi_f \mathbf{M}_e) + \nabla \cdot \left(-\mathbf{M}_e \frac{\kappa}{\eta}\nabla p - D\phi_f \nabla \mathbf{M}_e\right) = -\boldsymbol{\gamma}\phi_c\phi_f(\mathbf{M}_e - \mathbf{K}\mathbf{M}_i) \quad \text{in } \Omega, \tag{21a}$$

$$\left(-\mathbf{M}_e \frac{\kappa}{\eta}\nabla p - D\phi_f \nabla \mathbf{M}_e\right) \cdot \mathbf{n} = 0 \quad \text{at } \partial\Omega, \tag{21b}$$

where $\mathbf{M}_{e,i} = [A_{e,i} \ I_{e,i}]^T$, $\boldsymbol{\gamma} = \begin{bmatrix} \gamma_A^0 & 0 \\ 0 & \gamma_I^0 \end{bmatrix}$, and $\mathbf{K} = \begin{bmatrix} K_A & 0 \\ 0 & K_I \end{bmatrix}$. By observing Eq. 21 (a) & (b) and Eq. 7 (a) & (b), one can say that the intercellular forces dictate the spatial-temporal distribution of the active levels of extracellular morphogens. The active levels of the extracellular morphogens then influence those of the intracellular morphogens by endocytosis and exocytosis.

*2.4 The nondimensionalization of the governing equation systems*

To better interpret how model parameters influence the spatial-temporal distribution of morphogens, we nondimensionalize the governing equations Eq. 7 (a) & (b), Eq. 20 (a) & (b), and Eq. 21 (a)~(d) by the constants $\tau_A$ and $r$, which describe the time scale of chemical interactions between the activators and inhibitors, and the radius of the multicellular tissue, respectively. The nondimensionalized equation system is then written as:

$$\partial_{\bar{t}}(\phi_c \mathbf{M}_i) = \tau_A[\phi_c \mathbf{f}(A_i, I_i) + \boldsymbol{\gamma}\phi_c\phi_f(\mathbf{M}_e - \mathbf{K}\mathbf{M}_i)] \quad \text{in } \Omega, \tag{22a}$$

$$\partial_{\bar{t}}(\phi_f \mathbf{M}_e) - \frac{\tau_A D}{r^2}\overline{\nabla} \cdot (\mathbf{M}_e \overline{\nabla}\bar{p} + \phi_f \overline{\nabla}\mathbf{M}_e) = -\tau_A \boldsymbol{\gamma}\phi_c\phi_f(\mathbf{M}_e - \mathbf{K}\mathbf{M}_i) \quad \text{in } \Omega, \tag{22b}$$

$$(\mathbf{M}_e \overline{\nabla}\bar{p} + \phi_f \overline{\nabla}\mathbf{M}_e) \cdot \mathbf{n} = 0 \quad \text{at } \partial\Omega, \tag{22c}$$

$$\frac{\tau}{\tau_A}\partial_{\bar{t}}\phi_c + \bar{\nabla}\cdot\left(\frac{\tau D}{r^2}\bar{\nabla}\bar{p}\right) = \phi_h(\mathbf{M}_i) - \phi_c \quad \text{in } \Omega, \tag{22d}$$

$$\frac{\tau D}{r^2}\bar{\nabla}\bar{p}\cdot\mathbf{n} = 0 \quad \text{at } \partial\Omega. \tag{22e}$$

$$\bar{\nabla}\cdot\bar{\sigma} - \frac{\bar{Y}\phi_c}{\bar{h}_c}\bar{\mathbf{u}} = 0 \quad \text{in } \Omega, \tag{22f}$$

$$\bar{\sigma}\mathbf{n} = \mathbf{0} \quad \text{at } \partial\Omega, \tag{22g}$$

where $\partial_{\bar{t}}$ is the partial derivative with respect to the dimensionless time $\bar{t}$, while $\bar{\nabla}$ denotes for the gradient with respect to the dimensionless coordinates; $\mathbf{f}(A_i, I_i)$ stands for $[f(A_i, I_i) \; g(A_i, I_i)]^T$; $\bar{p}$ and $\bar{\sigma}$ are the dimensionless pressure and stress that is nondimensionalized by $D\eta/\kappa$; $\bar{\mathbf{u}}$ and $\bar{h}_c$ are dimensionless displacement and tissue thickness; and $\bar{Y} = Yr/(D\eta/\kappa)$ is the dimensionless tissue adhesion strength. Based on previous works, we can assume that $\tau/\tau_A \ll 1^{45,51}$, indicating that the cell volume relaxes infinitely faster than the chemical reactions of morphogens. As a result, Eq. (22d) is reduced to:

$$\bar{\nabla}\cdot\left(\frac{\tau D}{r^2}\bar{\nabla}\bar{p}\right) = \phi_h(\mathbf{M}_i) - \phi_c \quad \text{in } \Omega. \tag{23}$$

If $\phi_h$ and $\boldsymbol{\varepsilon}^A$ are insensitive to the active levels of morphogens, the spatial distribution of the $\phi_c$ will be independent on both $\bar{t}$ and $\mathbf{M}_i$, hence it can be decoupled from $\mathbf{M}_{e,i}$, which is shown in the next section.

## 3. Linear stability analysis of the weak-coupled model

To achieve mathematical tractability and perform linear stability analysis on the equation system of Eq. 22(a)~(g) and Eq. 23, it is possible to reduce the coupling of the model by assuming $k_{A,I} = 0$ in Eq. (6) and $\varepsilon^A$ to be a constant. Thereby, one can firstly solve the mechanical equilibrium coupled with Eq. (23) where $\phi_h = \phi^*$ under this assumption, and then substitute the solution of $\phi_c$ into Eq. 22(a)~(d) to solve for $\mathbf{M}_{e,i}$.

Let us firstly consider the tissue with no active contractility, which is $\varepsilon^{A*} = 0$. The cells occupy a homogeneous area fraction with a constant $\phi^*$ across the tissue. Under this condition, the solutions to the mechanical equilibrium are: $\bar{\mathbf{u}}^* = \mathbf{0}$, and $\bar{p}^* = 0$. The homeostatic solution of the active levels of morphogens can be readily derived as: $\mathbf{M}_{e,i}^* = [A_{e,i}^* \; I_{e,i}^*]^T$, where $A_e^* = K_A A_i^*$ and $I_e^* = K_I I_i^*$. Now we are interested in whether we can obtain a stationary spatial pattern for morphogen active levels if adding a small perturbation to the system. The perturbated active contractility then becomes $\varepsilon^A = \varepsilon^{A*} + \epsilon\delta\varepsilon^A$, where $\epsilon$ is a small positive number. Such perturbation causes the small change in the displacement field $\bar{\mathbf{u}} = \bar{\mathbf{u}}^* + \epsilon\delta\bar{\mathbf{u}}^*$ and the pressure $\bar{p} = \bar{p}^* + \epsilon\delta\bar{p}^*$, and finally leads to:

$$\mathbf{M}_{e,i} = \mathbf{M}_{e,i}^* + \epsilon\delta\mathbf{M}_{e,i}, \tag{24a}$$
$$\phi_c = \phi^* + \epsilon\delta\phi_c. \tag{24b}$$

By substituting Eq. 24(a) and (b) into Eq. 22(a)~(d) and noticing $\frac{\tau D}{l^2}\bar{\nabla}\cdot\bar{\nabla}\delta\bar{p} = -\delta\phi_c$ along with $\partial_{\bar{t}}\delta\phi_c = 0$, we can obtain th equations at the first order in $\epsilon$:

$$\phi^*\partial_{\bar{t}}\delta\mathbf{M}_i = \tau_A[\mathbf{f}_{\mathbf{M}_i^*}\phi^*\delta\mathbf{M}_i + \gamma\phi^*(1-\phi^*)(\delta\mathbf{M}_e - \mathbf{K}\delta\mathbf{M}_i)], \tag{25a}$$

$$(1-\phi^*)\partial_{\bar{t}}\delta\mathbf{M}_e + \frac{\tau_A}{\tau}\mathbf{M}_e^*\delta\phi_c - \frac{\tau_A D}{r^2}(1-\phi^*)\bar{\nabla}^2\delta\mathbf{M}_e = -\tau_A\gamma\phi^*(1-\phi^*)(\delta\mathbf{M}_e - \mathbf{K}\delta\mathbf{M}_i), \tag{25b}$$

where $\mathbf{f}_{\mathbf{M}_i^*} = \begin{bmatrix} f_{A_i^*} & f_{I_i^*} \\ g_{A_i^*} & g_{I_i^*} \end{bmatrix}$. As $\delta\phi_c$ now is independent on time, the solution to Eq. 25 (a) and (b) can be decomposed into a time-dependent general solution $\delta\mathbf{M}_{e,i}^0$ that satisfy the time-dependent equations:

$$\partial_{\bar{t}}\delta\mathbf{M}_i^0 = \tau_A[\mathbf{f}_{\mathbf{M}_i^*}\delta\mathbf{M}_i^0 + \gamma(1-\phi^*)(\delta\mathbf{M}_e^0 - \mathbf{K}\delta\mathbf{M}_i^0)], \tag{26a}$$

$$\partial_{\bar{t}}\delta\mathbf{M}_e^0 - \frac{\tau_A D}{r^2}\bar{\nabla}^2\delta\mathbf{M}_e^0 = -\tau_A\gamma\phi^*(\delta\mathbf{M}_e^0 - \mathbf{K}\delta\mathbf{M}_i^0); \tag{26b}$$

and a time-independent particular solution $\delta\mathbf{M}_{e,i}^1$ that satisfy the time-independent equations:

$$\mathbf{f}_{\mathbf{M}_i^*}\delta\mathbf{M}_i^1 + \gamma(1-\phi^*)(\delta\mathbf{M}_e^1 - \mathbf{K}\delta\mathbf{M}_i^1) = \mathbf{0}, \tag{27a}$$

$$\frac{1}{\tau}\mathbf{M}_e^*\delta\phi_c - \frac{D}{r^2}(1-\phi^*)\bar{\nabla}^2\delta\mathbf{M}_e^1 = -\gamma\phi^*(1-\phi^*)(\delta\mathbf{M}_e^1 - \mathbf{K}\delta\mathbf{M}_i^1), \tag{27b}$$

where $\delta\mathbf{M}_{e,i} = a\delta\mathbf{M}_{e,i}^0 + \delta\mathbf{M}_{e,i}^1$. Both $\delta\mathbf{M}_{e,i}^0$ and $\delta\mathbf{M}_{e,i}^1$ satisfy the boundary conditions and $a$ is solved by the initial condition. We next focus on solving Eq. 26 (a) and (b). We introduce the set of eigenvalues $\{\lambda_k^2\}_{k\geq 1}$ and the associated eigenvectors $\{U_k(\bar{\mathbf{x}})\}_{k\geq 1}$ which satisfy the following equation and boundary conditions:

$$\bar{\nabla}^2 U_k + \lambda_k^2 U_k = 0 \text{ in } \Omega, \tag{28a}$$
$$\bar{\nabla} U_k \cdot \mathbf{n} = 0 \text{ at } \partial\Omega. \tag{28b}$$

We then expand the general solution $\delta\mathbf{M}_{e,i}^0$ by the introduced sets of eigenvalues and eigenvectors[3]:

$$\delta\mathbf{M}_{e,i}^0(\mathbf{x},t) = \sum_{k=1}^{\infty}\mathbf{M}_{e,i}^{0k}U_k(\bar{\mathbf{x}})e^{\omega_k t}, \tag{29}$$

where $\omega_k$ is the growth rate with respect to the $k$th mode. For the one-dimensional case, $\lambda_k^2$ equals to $(\pi k)^2$, and $U_k(\bar{\mathbf{x}}) = \cos(\pi k\bar{x})$. While for a two-dimensional problem defined in a unit circle, $U_k(\bar{\mathbf{x}})$ can be expanded by a Fourier-Bessel series: $U_k(\bar{\mathbf{x}}) = J_{\alpha_k}(\lambda_k\bar{r})[a_{1_k}\cos(\alpha_k\theta) + a_{2_k}\sin(\alpha_k\theta)]$, where $\lambda_k$ is the $k$th positive root of $J'_{\alpha_k} = 0$. Due to the existence of non-zero solutions to Eq. 26 (a) and (b), we can derive that:

$$\begin{vmatrix} [\tilde{\omega}_k + \gamma_A^0 K_A(1-\phi^*) - f_{A_i^*}](\tilde{\omega}_k + \frac{D}{r^2}\lambda_k^2 + \gamma_A^0\phi^*) - (\gamma_A^0)^2 K_A(1-\phi^*)\phi^* & -f_{I_i^*}(\tilde{\omega}_k + \frac{D}{r^2}\lambda_k^2 + \gamma_I^0\phi^*) \\ -g_{A_i^*}(\tilde{\omega}_k + \frac{D}{r^2}\lambda_k^2 + \gamma_A^0\phi^*) & [\tilde{\omega}_k + \gamma_I^0 K_I(1-\phi^*) - g_{I_i^*}](\tilde{\omega}_k + \frac{D}{r^2}\lambda_k^2 + \gamma_I^0\phi^*) - (\gamma_I^0)^2 K_I(1-\phi^*)\phi^* \end{vmatrix}$$
$$=0, \tag{30}$$

where $\tilde{\omega}_k = \omega_k/\tau_A$. The solution should be stable for Eq. 26 (a) and (b) without the presence of diffusion, thus the fourth-order polynomial equation about $\omega_k$ in Eq. (30) should have all four roots with negative real parts when $D$ equals to zero. Such requirement leaves the coefficients of fourth-order polynomial all being positive:

$$(\gamma_A^0 K_A + \gamma_I^0 K_I)(1 - \phi^*) + (\gamma_A^0 + \gamma_I^0)\phi^* - tr(\mathbf{f_{M_i^*}}) > 0, \qquad (31\text{a})$$

$$[\phi^* \gamma_A^0 + (1-\phi^*)\gamma_A^0 K_A - f_{A_i^*}][\phi^* \gamma_I^0 + (1-\phi^*)\gamma_I^0 K_I - g_{I_i^*}] - \phi^*(g_{I_i^*}\gamma_I^0 + f_{A_i^*}\gamma_A^0) - f_{I_i^*}g_{A_i^*} > 0, \qquad (31\text{b})$$

$$-f_{A_i^*}\gamma_A^0[\phi^* \gamma_I^0 + (1-\phi^*)\gamma_I^0 K_I] - g_{I_i^*}\gamma_I^0[\phi^* \gamma_A^0 + (1-\phi^*)\gamma_A^0 K_A] + (\gamma_A^0 + \gamma_I^0)\det(\mathbf{f_{M_i^*}}) > 0, \qquad (31\text{c})$$

$$\gamma_A^0 \gamma_I^0 \phi^{*2} \det(\mathbf{f_{M_i^*}}) > 0. \qquad (31\text{d})$$

inhibitor system require that $tr(\mathbf{f_{M_i^*}}) < 0$ and $\det(\mathbf{f_{M_i^*}}) > 0$.[38,45] We point out here that such requirement can fulfill the above inequalities if $K_A, K_I \ll 1$, and $\gamma_A K_A$, $\gamma_I K_I$ are in the same order of magnitude.

**4. Choice of model parameters**

In this section, we discuss how the model parameters are determined. Firstly, the colony radius $r$ used to nondimensionalize the model is set to be $r \sim 10^2$ μm, which corresponds to a typical pattern size for the two-dimensional developmental hPSC tissue[24]. While the time scale $\tau_A$ that describes the time scale of bio-chemical interactions of morphogens has been measured as $\tau_A \sim 10^4 - 10^5$ s.[61] The volume relaxation time scale $\tau$ is estimated as $\tau \sim 10^2$ s which agrees with the time scale involved in volume regulation of cells under an osmotic perturbation[45,51]. Thus, the assumption $\tau/\tau_A \ll 1$ holds for Eq. (23). The cellular phase at the homogeneous stationary state is assumed to be: $\phi^* \cong 1$, which indicates that the cells fulfill the tissue[45]. The parameters $k_A$ and $k_I$ in Eq. (6) are estimated in the way that $\phi_c$ does not vary dramatically. Next, we estimate the model parameters related with the endocytosis and exocytosis of the cells and the parameters related to the chemical reactions of the two morphogens. From a biological standpoint, the effective endocytosis rates satisfy $\gamma_{A,I}^0 \phi_c \phi_f \sim 10^{-2}$ s$^{-1}$,[62] thus $\gamma_{A,I}^0 \sim 10^0 - 10^{-1}$ s$^{-1}$ as $\phi_c \phi_f \sim 10^{-2}$. The coefficients of the biochemical reaction rates should be around: $f_{A_i^*,I_i^*}$ & $g_{A_i^*,I_i^*} \sim 10^{-4}$ s$^{-1}$.[45] The endocytosis-exocytosis-rate ratios $K_A$ and $K_I$ are much less than 1, hence are set to be $K_{A,I} \sim 10^{-2} - 10^{-1}$ and $K_A < K_I$ according to the previous study[45]. The parameters for the interstitial fluid are estimated in the following way. The global Fickian diffusivity $D = D_0$ is in the order of magnitude of $10^{-13} - 10^{-8}$ m$^2 \cdot$s$^{-1}$;[45] the tissue permeability $\kappa$ should be $\kappa \sim 10^{-16} - 10^{-20}$ m$^2$;[57] the viscosity is $\eta \approx 10^{-3}$ Pa$\cdot$s as the interstitial fluid is mainly consisted of water. At last, the mechanical properties of the developmental porous tissue are estimated in the following way: the shear modulus $G$ is estimated as $G \sim 10^3$ Pa;[63] by considering the Poisson's ratio of cells is usually from 0.3 to 0.5,[27,64,65] the drained and undrained Poisson's ratios of the cellular tissue are estimated in the range of $0.4 - 0.5$,[41,66,67] and should be very close with each other[41,66]. As a result, the drained bulk modulus $K$ is approximately $K \sim 10^4$ Pa, which corresponds to previous experimental measurements[68]. For biological tissues, Biot coefficient $\alpha$ is close to 1.[45,67,69] The active contractility is chosen to be $\varepsilon^A \sim 10^{-2}$ so that the corresponding active stress $\sigma^A = \frac{2G(1+\nu_u)}{1-\nu_u}\varepsilon^A$ is in the order of $10^2 - 10^3$ Pa.[27,64] The effective adhesion strength of focal adhesions $Y$ is set to be $Y \sim 10^8$ N·m$^{-3}$. The tissue thickness is chosen to be $h_c \sim 10$ μm,[27] which corresponds to $\bar{h}_c \sim 10^{-1} - 10^{-2}$. The detailed number of these parameters are listed in Tab. 1.

**Table 1**
Model parameters adopted in this article

| Symbol | Definition | Value |
|---|---|---|
| $r$ | Radius of the tissue | $2 \times 10^2$ μm |
| $\tau_A$ | Characteristic time scale of the chemical reactions of morphogens | $10^4$ s |
| $\tau$ | Time scale of the cell volume regulation | $10^2$ s |
| $\phi^*$ | Cellular phase at the homogeneous stationary state | 0.9 |
| $k_A$ | Sensitivity of the cell volume to the active level of the activators | 1.5 |
| $k_I$ | Sensitivity of the cell volume to the active level of the inhibitors | 3 |
| $\gamma_{A,I}^0$ | Endocytosis rate of the activators and inhibitors | $\frac{1}{9}$ s$^{-1}$ |
| $\begin{bmatrix} f_{A_i^*} & f_{I_i^*} \\ g_{A_i^*} & g_{I_i^*} \end{bmatrix}$ | Chemical-reaction-rate-related parameters | $\begin{bmatrix} 9.989 & 11.11 \\ -9.989 & -10.11 \end{bmatrix} \times 10^{-4}$ s$^{-1}$ |
| $f_{A_i^* I_i^* I_i^*}$ | Parameter that describes the inhibition made by inhibitors acting on activators | $-3.15 \times 10^{-2}$ s$^{-1}$ |
| $g_{A_i^* I_i^* I_i^*}$ | Parameter that describes the promotion made by activators acting on inhibitors | $3.15 \times 10^{-2}$ s$^{-1}$ |
| $K_A$ | Ratio between the endocytosis and exocytosis rates with respect to the activators | 0.1 |
| $K_I$ | Ratio between the endocytosis and exocytosis rates with respect to the inhibitors | 0.1938 |
| $D$ | Morphogen diffusivity in the interstitial fluid | $5.9644 \times 10^{-12}$ m$^2$·s$^{-1}$ |
| $\kappa$ | Tissue permeability | $3.5870 \times 10^{-18}$ m$^2$ |
| $\eta$ | Viscosity of the interstitial fluid | $10^{-3}$ Pa·s |
| $K$ | Drained bulk modulus of the | $1.6667 \times 10^4$ Pa |

|   |   | tissue |   |
|---|---|---|---|
| $\nu$ | Drained Poisson's ratio |   | 0.4 |
| $\nu_u$ | Undrained Poisson's ratio |   | 0.42 |
| $\alpha$ | Biot coefficient |   | 0.8 |
| $\beta$ | Material parameter that relates the fluid phase and the active contraction |   | 0.4 |
| $\varepsilon_0^A$ | Active cell contractility in cytoskeleton |   | 0.048 |
| $Y$ | Effective adhesion strength of focal adhesions |   | $2 \times 10^8$ N·m$^{-3}$ |
| $h_c$ | Tissue thickness |   | 10 μm |

## 5. Numerical simulations and discussions

In this section, we solve the equations in Section 2 with the giving parameters in Section 4 numerically by finite elements method implanted in **COMSOL Multiphysics** software. The initial conditions for the intracellular active level of both morphogens $A_i^0$ and $I_i^0$ are set to be random, ranging from $3.75 \times 10^{-3}$ to $6.25 \times 10^{-3}$ whose averages are $A_i^* = I_i^* = 5 \times 10^{-3}$. The initial conditions for the extracellular active level of both morphogens, on the other hand, are set to be: $A_e^0 = K_A A_i^0$ and $I_e^0 = K_I I_i^0$. These settings aim to mimic the small perturbations existing in the active levels of morphogens at the beginning of the reaction-diffusion process. This section is arranged in the following way: Section 5.1 studies the weakly-coupled model where $\phi_h$ and $\varepsilon^A$ are constant; Section 5.2 ~ 5.6 study the fully-coupled model where $\phi_h$ and $\varepsilon^A$ are functions on $A_i$ and $I_i$.

*5.1 Predictions of ring pattern formation in circular-shaped developmental tissue by the weakly-coupled model*

In the weakly-coupled model, the parameters $k_{A,I} = 0$, thus Eq. (6) reduces to $\phi_h = \phi^*$, indicating that the cell volume is insensitive to the presence of the morphogens. The active contraction $\varepsilon^A$ equals to $\varepsilon_0^A$, which is also a constant. As a result, the stress equilibrium and the evolution of the cellular phase are time and morphogen independent. The predictions on the cellular phase and the distribution of the active level of the morphogens are presented in Fig. 2 at the steady state.

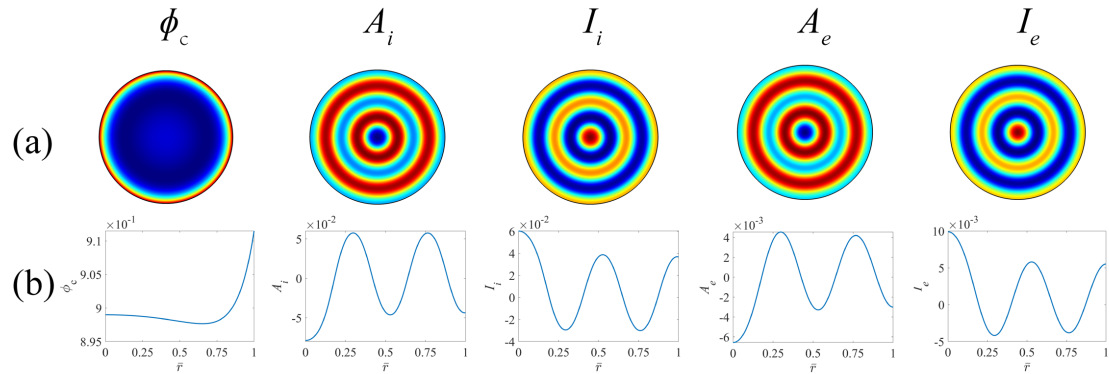

**Fig. 2.** The contour and curve plot of the predictions made by the weakly-coupled model for the circular tissues. (a) The contour plot of the distribution of the cellular phase variable $\phi_c$ and the distribution of the active levels of intracellular and extracellular morphogens: $A_i$, $I_i$, $A_e$, and $I_e$ (red for large/positive values, and blue for small/negative values). (b) The curve plot of the cellular phase variable $\phi_c$ and the distribution of the active levels of intracellular and extracellular morphogens along the radius of the tissue.

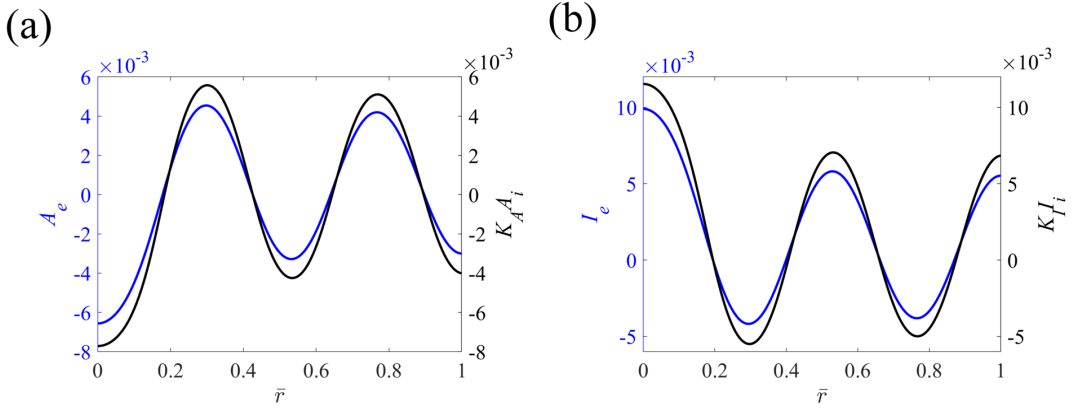

**Fig. 3.** (a) The curve plot of $K_A A_i$ and $A_e$ along the radius of the circular tissue derived by the weakly-coupled model. (b) The curve plot of $K_I I_i$ and $I_e$ along the radius of the circular tissue by the weakly-coupled model.

The results plotted in Fig. 2 show axisymmetric pattens instead of random stripes and dots when mechanical forces are not taken into considerations[11,19]. Such results imply a strong coupling between the mechanical forces and the reaction-diffusion of the morphogens. By observing the first column in Fig. 2, one can see that the cellular phase variable $\phi_c$ firstly decreases from the center of the tissue and then increases monotonically until it reaches its maximum at the boundary. Such predictions show that the cells compact more closely at the tissue boundary than they do in the intermediate region. Additionally, one could see that the variation in $\phi_c$ across the whole tissue is relatively small, which agrees with the fact that previous studies did not report an obvious change in the cellular area fraction across the patterned developmental tissue.[24,25] And it also satisfies the assumption made in Section 2 that the variation in $\phi_c$ is not large enough to affect the global Fickian diffusivity. The second to the fifth columns in Fig 2. show the spatial distributions of the active levels of the activators and inhibitors inside and outside cells. The spatial distributions of these active levels are axisymmetric and form layers of concentric rings that resemble the pattern of wave propagations with peaks and troughs. When the active levels of the activators reach the maximum, those of the inhibitors reach the minimum, and vice versa. We point out that the predictions made by the weakly-coupled model agree well with the previous experiments[22] on the distributions of active levels of morphogens. Another conclusion that we could draw from Fig. 2 is that the distribution of the active levels of the intracellular and extracellular morphogens are highly corelated. In details, the positions of the peaks and troughs of $A_i$ are roughly the same with those of $A_e$, so do $I_i$ and

$I_e$. Next, we ask how the active levels of the intracellular and extracellular morphogens are correlated with each other by plotting the values of $K_A A_i$ versus $A_e$ and $K_I I_i$ versus $I_e$ along the tissue radius in Fig 3. The results show that the landscapes of $K_A A_i$ and $K_I I$ are very similar with those of $A_e$ and $I_e$, respectively. Moreover, the values of $K_A A_i$ and $K_I I$ are also scaled with those of $A_e$ and $I_e$, with a 40% relative error in maximum.

The landscapes of the traction forces $|\mathbf{T}| = Y\phi_c|\mathbf{u}|$ and the intercellular tension $\sigma = \frac{\sigma_{11}+\sigma_{22}}{2}$ are plotted in Fig. 4. The magnitude of the traction force gradually increases along the radial direction until it reaches the maximum at the tissue boundary.[70] The intercellular tension, on the other hand, firstly ramps up in the central region of the tissue and then decreases to its minimum at the tissue boundary.[70] We note that the predictions made here resemble the predictions where $\phi_c = const$ as the variations in $\phi_c$ is small.

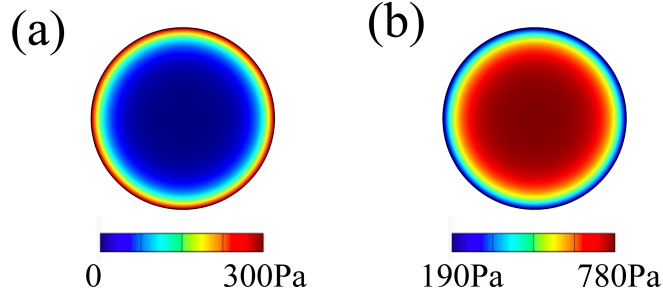

**Fig. 4.** The landscapes of the traction forces and the intercellular tension predicted by the weakly-coupled model. (a) The landscape of the traction forces. (b) The landscape the intercellular tension.

*5.2 Predictions of ring pattern formation in circular-shaped developmental tissue by the fully-coupled model*

In the fully-coupled model, the parameters $k_{A,I}$ are set to be $k_A = 1.5$ and $k_I = 3$, as the cell volume is assumed to be more sensitive to the presence of the inhibitors than the activators, suggested by Recho et al[45]. As another supplement to the weakly-coupled model, we assume that the active contraction $\varepsilon^A$ is also affected by the active levels of the morphogens. Although it is usually considered being a function of substrate stiffness[27,70], the active contractility can be affected by morphogens that act as growth factors. It is found that the active contraction can either be increased under the exposure of IGF or EGF in virto[71,72], or be suppressed or dysfunctional by FGF[73]. We assume that such effect can be written in a linear form based on the previous work[44]:

$$\varepsilon^A = \varepsilon_0^A[1 + k_\varepsilon(A_i - I_i)], \tag{32}$$

The value of $k_\varepsilon$ is set to be $-3$, which indicates that the activators and inhibitors are able to weaken and strengthen the active contraction, respectively. The predictions on the cellular phase and the distribution of the active level of the morphogens are presented in Fig. 5 at the steady state.

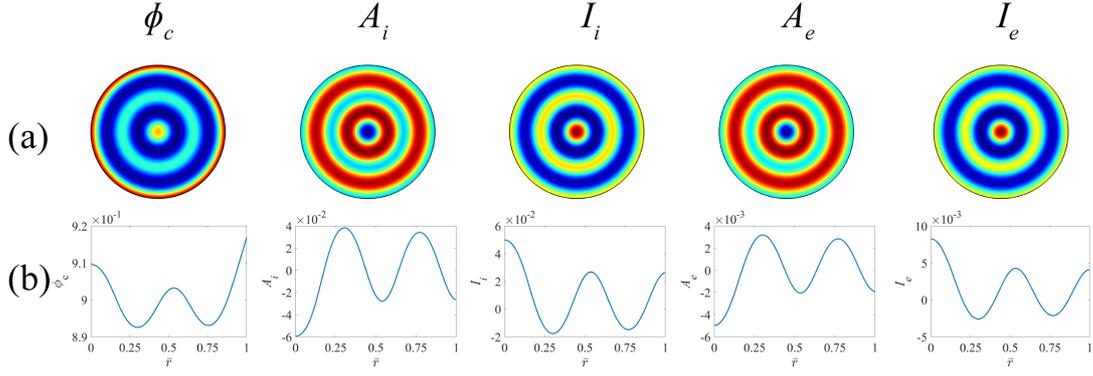

**Fig. 5.** The contour and curve plot of the predictions made by the fully-coupled model for the circular tissue. (a) The contour plot of the distribution of the cellular phase variable $\phi_c$ and the distribution of the active levels of intracellular and extracellular morphogens: $A_i$, $I_i$, $A_e$, and $I_e$ (red for large/positive values, and blue for small/negative values). (b) The curve plot of the cellular phase variable $\phi_c$ and the distribution of the active levels of intracellular and extracellular morphogens along the radius of the tissue.

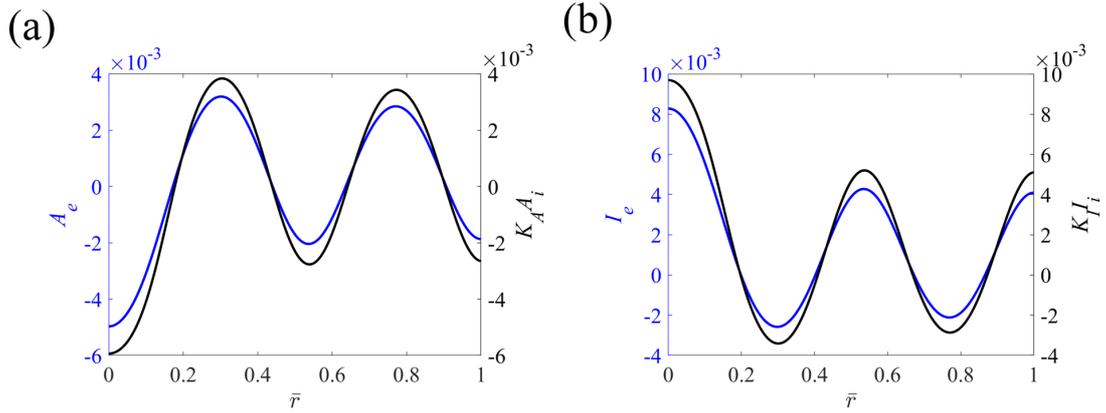

**Fig. 6.** (a) The curve plot of $K_A A_i$ and $A_e$ along the radius of the circular tissue derived by the fully-coupled model. (b) The curve plot of $K_I I_i$ and $I_e$ along the radius of the circular tissue by the fully-coupled model.

By observing the first column in Fig. 5, one can see that the cell area fraction $\phi_c$ now forms a concentric ring pattern similar to that of the morphogens. The distribution curve of $\phi_c$ wiggles in the centroid region of the tissue and then increases monotonically and reaches its maximum at the boundary. Yet still, the variation in $\phi_c$ across the whole tissue remains roughly small, which does not violate the assumption made in Section 2 so that the global Fickian diffusivity $D$ can still be considered as a constant. The spatial distributions of the active levels of the morphogens highly resemble those predicted by the weakly-coupled model. However, by comparing Fig. 5(b) with Fig. 3(b), one can tell that the variations between the peaks and troughs predicted by the fully-coupled model are a little smaller than those by the weakly-coupled model. And the correlations of $A_e \sim K_A A_i$, and $I_e \sim K_I I_i$ also hold for the fully-coupled model, as shown in Fig. 6.

As hypothesized in Eq. (32), the active levels of morphogens can shape the landscapes of the traction forces and the intercellular tension. As plotted in Fig. 7(a), the magnitude of the traction force fluctuates within the central region of the tissue and then reaches the maximum at the tissue boundary. In Fig. 7 (b), the landscape of the intercellular tension also fluctuates in the central region and gradually reaches its minimum at the tissue boundary, which is quite unlike the predictions made by the weakly-coupled model in Fig. 4.

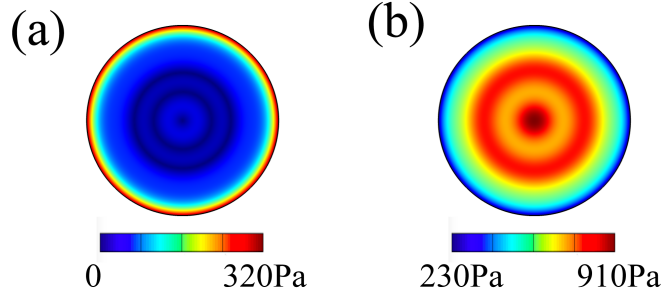

**Fig. 7.** The active levels of morphogens can shape the landscapes of the traction forces and the intercellular tension predicted by the fully-coupled model. (a) The landscape of the traction forces. (b) The landscape the intercellular tension.

*5.3 Predictions of pattern formations in developmental tissue with different values of the global Fickian diffusivity by the fully-coupled model*

Previous studies show that the pattern formation of morphogens is also dictated by the global Fickian diffusivity $D$.[13,19] Although not being a function of tissue deformation or the cellular phase variable $\phi_c$, $D$ still can be affected by the tissue tortuosity, created by cells as obstacles that increase the diffusion path length for the morphogens. By changing either the cell number density or the cellular geometry, the tissue tortuosity can be significantly modified without changing $\phi_c$ and leads to great variations in the value of the global Fickian diffusivity $D$.[13] In Fig. 8, we plot the predictions made by different values of $D$, where $D_0$ takes the value in Tab. 1, $D_1 = \frac{16}{9} D_0$, and $D_2 = 4D_0$.

Different values of the diffusivity $D$ can lead to different numbers of peaks and troughs in the active levels of morphogens. As one can see in Fig. 8, $D_0$ leads to two peaks and one trough in the active levels of the activators, and two troughs and one peak in the active levels of the inhibitors, respectively. Yet, $D_1$ gives one peak and one trough for both the activators and inhibitors. The diffusivity $D_2$ can only lead to one peak for the activator and one trough for the inhibitor. The trend is clear that the increasing diffusivity decreases the number of the fluctuation period in the active levels of both morphogens. Additionally, $D_0$ and $D_2$ both predict high active levels for the inhibitors and low active levels for the activators in the center of the tissue. Differing from that, $D_1$ predicts a reversed result that shows low active levels for the inhibitors and high active levels for the activators in the center of the tissue. It indicates that the change in the value of the diffusivity may lead to a change in the signs of the active levels of both morphogens.

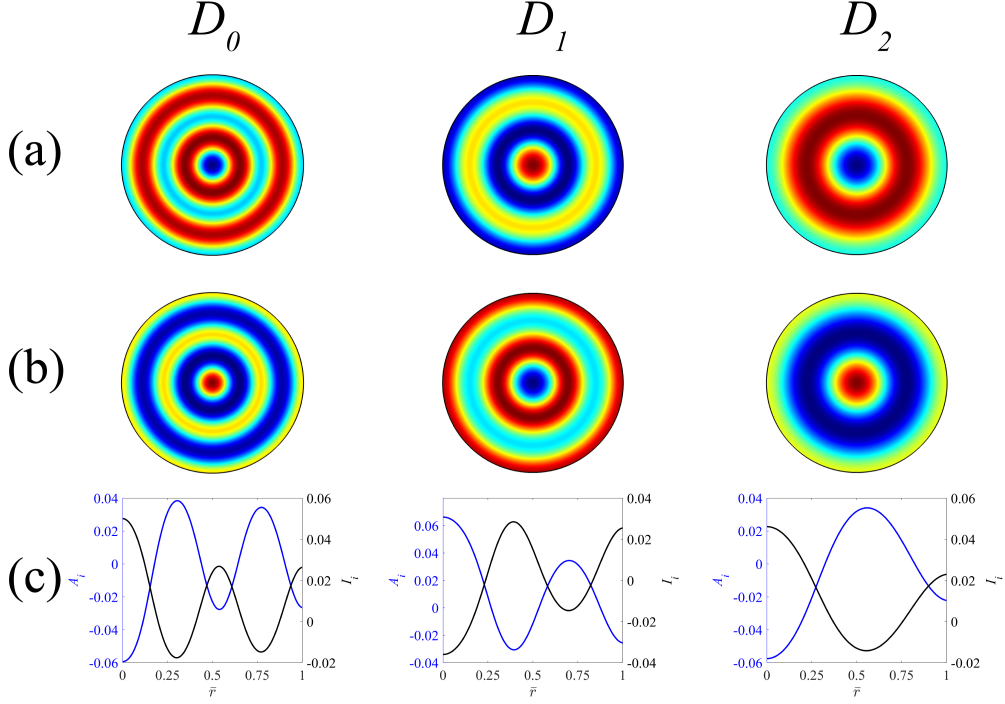

**Fig. 8.** The contour and curve plot of the predictions made by the fully-coupled model for the circular tissues with respect to different diffusivities: $D_0$, $D_1$ and $D_2$. (a) The contour plot of the distribution of the active levels of intracellular morphogens: $A_i$ (red for large/positive values, and blue for small/negative values). (b) The contour plot of the distribution of the active levels of intracellular morphogens: $I_i$. (c) The curve plot of the distribution of the active levels of intracellular and extracellular morphogens along the radius of the tissue.

*5.4 Predictions of pattern formations in developmental tissue with different endocytosis and exocytosis rates with respect to the activators and inhibitors by the fully-coupled model*

In this section, we study how different endocytosis/exocytosis rates with respect to the two kinds of morphogens can affect the pattern formation results. Firstly, we choose $\gamma_A^0 = \frac{2}{9}$ s$^{-1}$, and let $\gamma_I^0 = \frac{1}{9}$ s$^{-1}$ remain unchanged, which means the endocytosis/exocytosis of the activator is two times faster than that of the inhibitor. The pattern formation and the magnitude of the active levels change significantly, as plotted in the second column of Fig. 9. Specifically speaking, the amplitude of the morphogen active levels reduces from the order of $10^{-2}$ to $10^{-3}$. The amplitude of the active levels in the central region is much lower than that in the tissue periphery. Next, we let $\gamma_I^0$ equal to $\frac{2}{9}$ s$^{-1}$, and $\gamma_A^0$ remain unchanged, meaning the endocytosis/exocytosis of the inhibitor is now two-fold faster. The predicted results show that the amplitude of the morphogen active levels increases from the order of $10^{-2}$ to $10^{-1}$. And the peaks and troughs in the active levels of either the activators or the inhibitors have almost identical absolute values.

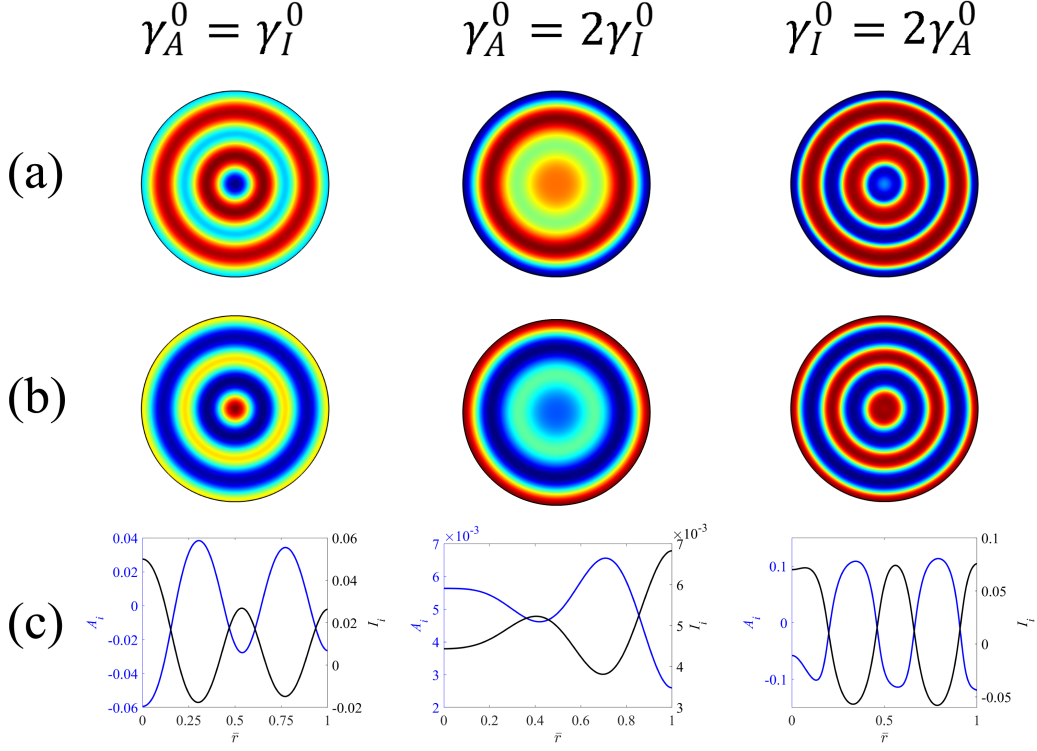

**Fig. 9.** The contour and curve plot of the predictions made by the fully-coupled model for the circular tissues with respect to different combinations of the values of $\gamma_{A,I}$, from the left to the right: $\gamma_A^0 = \gamma_I^0 = \gamma_0$, $\gamma_A^0 = 2\gamma_I^0 = 2\gamma_0$, and $\gamma_I^0 = 2\gamma_A^0 = 2\gamma_0$. (a) The contour plot of the distribution of the active levels of intracellular morphogens: $A_i$ (red for large/positive values, and blue for small/negative values). (b) The contour plot of the distribution of the active levels of intracellular morphogens: $I_i$. (c) The curve plot of the distribution of the active levels of intracellular and extracellular morphogens along the radius of the tissue.

*5.5 Predictions of pattern formations in developmental tissue with different tissue sizes by the fully-coupled model*

In Section 5.5, we examine how the tissue size can affect the landscapes of the active levels of the morphogens by studying the three cases where the dimensionless radius of circular tissues equals $\bar{R} = 0.5$, $\bar{R} = 1$, and $\bar{R} = 2$, respectively. The predicted results are plotted in Fig. 10. By looking into Fig. 10, we can observe that, from the left to the right column, the numbers of the peaks and troughs of the morphogens gradually increase as the dimensionless radius increases from 0.5 to 2. To be specific, for $\bar{R} = 0.5$, there is only one peak in the active level of the activator and one trough in that of the inhibitor. For $\bar{R} = 1$, the number of the peaks in the active level of the activator increases to 2 and so does the number of the troughs in the active level of the inhibitor. For $\bar{R} = 2$, the numbers both become 4, indicating that they are scaled with the tissue radius by a proportion of 2. By recalling the relation $A_e \sim K_A A_i$, and $I_e \sim K_I I_i$ derived in Section 5.1 and 5.2, we can sum Eq. (22a) and (22b) together:

$$\partial_{\bar{t}}[(\phi_c + \phi_f \mathbf{K})\mathbf{M}_i] - \frac{\tau_A D}{l^2} \bar{\nabla} \cdot \mathbf{K}(\mathbf{M}_i \bar{\nabla}\bar{p} + \phi_f \bar{\nabla}\mathbf{M}_i) \approx \tau_A \phi_c \mathbf{f}(A_i, I_i), \qquad (33)$$

Based on above equation and previous studies[25], the general Turing instability in a unit circle should lead to a scaling law, saying that:

$$\frac{D(1-\phi^*)K_A}{l^2} \sim \frac{(f_{A_i^*}+g_{I_i^*}h)\phi^*}{2k^2}, \quad (34)$$

where $h$ is estimated as $h \sim K_A/K_I$. The wavenumber $k$ then can be calculated by the coefficients adopted in Tab. 1: $k \approx 12$. The numbers of the peaks and troughs are then equal to $\frac{k}{2\pi} \approx 2$, which corresponds to the scaling proportion we derived above.

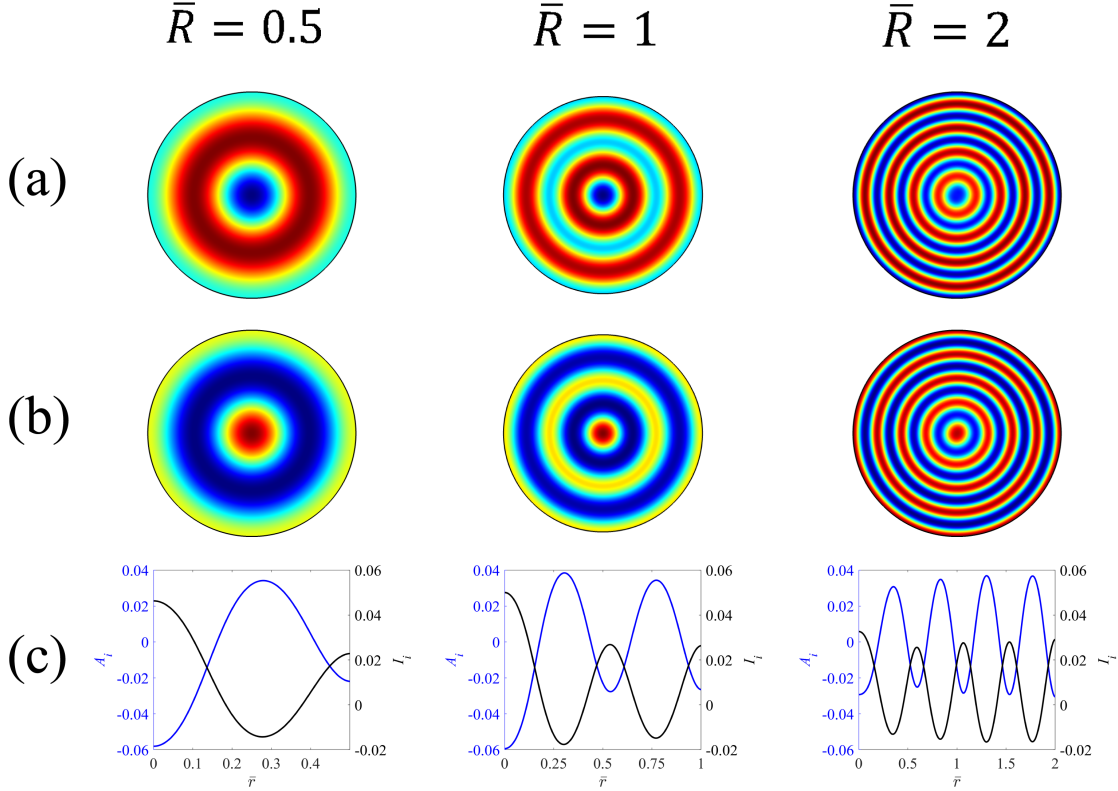

**Fig. 10.** The contour and curve plot of the predictions made by the fully-coupled model for the circular tissues with respect to different dimensionless radius $\bar{R}$, from the left to the right: $\bar{R} = 0.5$, $\bar{R} = 1$, and $\bar{R} = 2$. (a) The contour plot of the distribution of the active levels of intracellular morphogens: $A_i$ (red for large/positive values, and blue for small/negative values). (b) The contour plot of the distribution of the active levels of intracellular morphogens: $I_i$. (c) The curve plot of the distribution of the active levels of intracellular and extracellular morphogens along the radius of the tissue.

*5.6 Predictions of pattern formations in developmental tissue with different geometries by the fully-coupled model*

At last, we examine how geometry affects the pattens formed by the active levels of morphogens. In Fig. 11, we apply the fully-coupled model onto the equilateral triangular-shaped and square-shaped tissues, where the predicted results imply that they are highly regulated by domain geometries. The active levels of the morphogens, as well as the cellular phase variable $\phi_c$ show a 120-degree rotational symmetry for the triangular-shaped tissue as plotted in Fig. 11(a). For the square-shaped tissue, the active levels of the morphogens and the phase variable $\phi_c$ show a 90-degree rotational symmetry. The symmetries of these variables are identical with the symmetries of the shape of the tissues as the distributions of the morphogens are dictated by the pressure within the tissues. The correlations between the active levels of the intracellular and extracellular morphogens, as plotted in the last two columns of Fig. 11, still holds for tissues with different shapes.

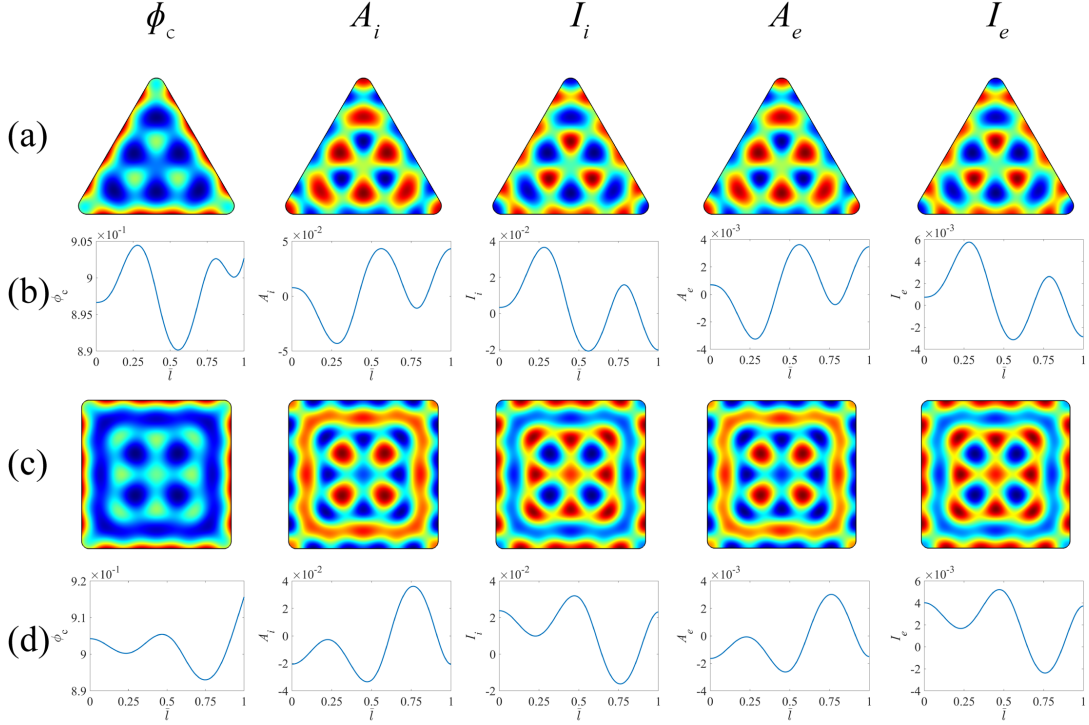

**Fig. 11.** The contour and curve plot of the predictions made by the fully-coupled model for the tissues with different geometries. (a) The contour plot of the distribution of the cellular phase variable $\phi_c$ and the distribution of the active levels of intracellular and extracellular morphogens: $A_i$, $I_i$, $A_e$, and $I_e$ for the equilateral-triangular-shaped tissue (red for large/positive values, and blue for small/negative values), with an edge-length and the radius of the rounded corner equal to 2 and 0.1, respectively. (b) The curve plot of the cellular phase variable $\phi_c$ and the distribution of the active levels of intracellular and extracellular morphogens of the triangular-shaped tissue along the line which points from the tissue centroid to one of the tissue corners. (c) The contour plot of the distribution of the cellular phase variable $\phi_c$ and the distribution of the active levels of intracellular and extracellular morphogens: $A_i$, $I_i$, $A_e$, and $I_e$ for the square-shaped tissue (red for large/positive values, and blue for small/negative values), with an edge-length and the radius of the rounded corner equal to 2 and 0.1, respectively. (d) The curve plot of the cellular phase variable $\phi_c$ and the distribution of the active levels of intracellular and extracellular morphogens of the square-shaped tissue along the line which starts from the tissue centroid and aligns with the in-plane horizontal axis.

## 6. Conclusions

We raise up a biphasic contraction-reaction-diffusion poroelastic model to understand the pattern formation of the activators and inhibitors in two-dimensional geometrically confined hPSCs-derived microtissues during its stage of cell differentiation. The model predicts a distribution of wave-like active levels of the morphogens with random initial conditions. Such distribution, in turn, shapes the evolutions of the cellular phase, the tissue active contractility, and the landscapes of the intercellular and the extracellular forces. The active levels of the intracellular and extracellular morphogens are highly corelated, revealed by our simulation. The number of the peaks and troughs in the distribution of the active levels of both morphogens are regulated by the global Fickian diffusivity. We also find that the rates of the endocytosis and exocytosis of the morphogens can significantly affect the pattern formation of the morphogens by both changing the amplitude of the active levels and the mode of the pattern formation. Next, we examine the distribution of morphogen active levels for circular tissues with different sizes. The results show that the numbers of peaks and troughs in the morphogens active levels are proportional to the tissue radius. Finally, we claim that the pattern formation is dictated by the geometry of the confined tissues. For the circular-shaped tissue, the pattern formation is usually axisymmetric; for the equilateral triangular-shaped ones, it is 120-degree rotational symmetric; and for the square-shaped ones, the pattern forms a 90-degree rotational symmetry. We hope that the proposed model can shed light on the study of biophysical mechanisms of early-stage pattern formation during embryogenesis, especially in the hPSCs-derived in vitro models.

## 7. Declaration of competing interest
The authors declare that they have no competing financial interest in the work reported in this paper.

## 8. Acknowledgement

H.Y. acknowledge the funding support from the Southern University of Science and Technology, China (sustech.edu.cn).

Growth Factor-23 Promotes Rhythm Alterations and Contractile Dysfunction in Adult Ventricular Cardiomyocytes. *Nephrol. Dial. Transplant.* **2019**, *34* (11), 1864–1875. https://doi.org/10.1093/ndt/gfy392.